# Hydrogen from Cellulose and Low-density Polyethylene via Atmospheric Pressure Nonthermal Plasma

**Benard Tabu**[1], **Visal Veng**[1], **Heba Morgan**[2], **Shubhra Kanti Das**[1], **Eric Brack**[3], **Todd Alexander**[3], **J. Hunter Mack**[1], **Hsi-Wu Wong**[2], **and Juan Pablo Trelles**[1*]

[1] Department of Mechanical Engineering, University of Massachusetts Lowell, United States of America
[2] Department of Chemical Engineering, University of Massachusetts Lowell, United States of America
[3] US Army Combat Capabilities Development Command Soldier Center, United States of America
[*] Corresponding author, Juan_Trelles@uml.edu

**Abstract**

The valorization of waste, by creating economic value while limiting environmental impact, can have an essential role in sustainable development. Particularly, polymeric waste such as biomass and plastics can be used for the production of green hydrogen as a carbon-free energy carrier through the use of nonthermal plasma powered by renewable, potentially surplus, electricity. In this study, a Streamer Dielectric-Barrier Discharge (SDBD) reactor is designed and built to extract hydrogen and carbon co-products from cellulose and low-density polyethylene (LDPE) as model feedstocks of biomass and plastic waste, respectively. Spectroscopic and electrical diagnostics, together with modeling, are used to estimate representative plasma properties, namely electron and excitation temperatures, number density, and power consumption. Cellulose and LDPE are plasma-treated for different treatment times to characterize the evolution of the hydrogen production process. Gas products are analyzed using gas chromatography to determine the mean hydrogen production rate, production efficiency, hydrogen yield, selectivity, and energy cost. The results show that the maximum hydrogen production efficiency for cellulose is 0.8 mol/kWh, which is approximately double that for LDPE. Furthermore, the energy cost of hydrogen production from cellulose is 600 kWh/kg of $H_2$, half that of LDPE. Solid products are examined via scanning electron microscopy, revealing the distinct morphological structure of the two feedstocks treated, as well as by elemental composition analysis. The results demonstrate that SDBD plasma is effective at producing hydrogen from cellulose and LDPE at near atmospheric pressure and relatively low-temperature conditions in rapid-response and compact processes.

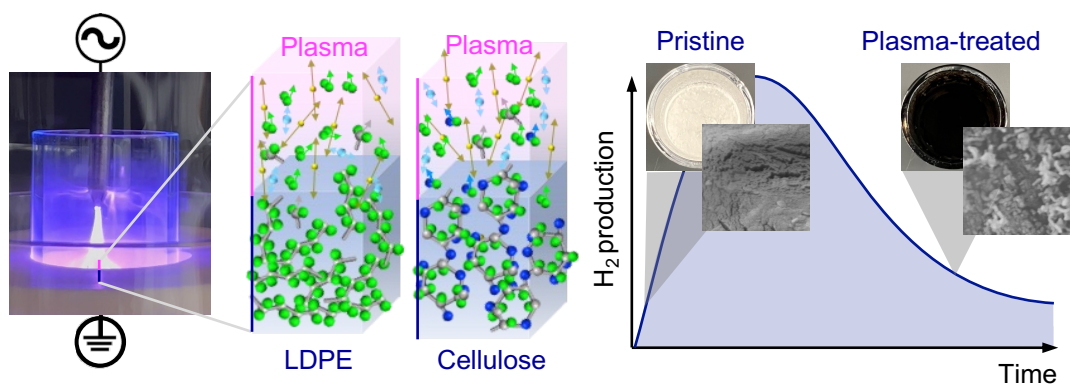

# 1. Introduction

The utilization of fossil-based resources is the major contributor to greenhouse gas emissions leading to environmental pollution and climate change. According to Li [1], 31.5 Gt of $CO_2$ was generated in 2022 despite the low economic activities attributed to the COVID-19 pandemic. Such substantive $CO_2$ emissions are mainly due to the use of fossil fuels, accounting for over 80% of global energy consumption [2]–[5]. The use of alternative fuels, particularly green hydrogen (i.e., hydrogen generated using renewable energy [6]) derived from non-fossil feedstock such as plastic and biomass waste, can support the creation of economic value while limiting environmental impacts. The increasing amount of organic polymeric waste, estimated to reach 25 Gt of plastic waste globally by 2050 [7] and 146 billion metric tons of biomass annually [8], [9], can be considered an enormous resource, particularly for the production of hydrogen via processes powered by renewable, potentially surplus, electricity.

Methane-steam reforming and water electrolysis are currently the dominant methods for the production of hydrogen. Water electrolysis is the primary approach for carbon-free hydrogen production (when powered by renewable electricity). Nevertheless, due to its relatively high energy cost compared to methane-steam reforming [10], electrolytic approaches account for only 4% of total hydrogen production [11]. Other environmentally benign hydrogen production methods in development include photocatalytic, photobiological, and photochemical water splitting [12]–[16].

In contrast to the use of methane or water as feedstock, the use of solids, particularly polymeric waste, for hydrogen production remains largely untapped. Aziz *et al.* [17] noted that converting biomass and other organic solid materials to hydrogen is a promising approach due to feedstock availability and could lead to positive economic, social, and environmental impacts. The main methods for the production of hydrogen from solids are pyrolysis and gasification [18], [19]. These methods can be thermo-chemical (using temperature and pressure as the main process parameters), thermo-catalytic (incorporating catalysts), or thermal plasma-based, and generally focus on the production of syngas, a mixture of mainly hydrogen and carbon monoxide. Thermal plasma methods are particularly appealing because they make direct use of electricity, which given the increasing capacity of renewable electricity generation and limited electricity storage, is sometimes available as surplus. Moreover, thermal plasma methods are robust for the treatment of heterogeneous and hard-to-decompose waste streams, require minimal or no consumables, and do not rely on catalysts. Nevertheless, pyrolysis and gasification processes – including those based on thermal plasma – generally operate with low energy efficiency (typically defined as the caloric content of syngas produced per unit energy consumed) and/or low selectivity towards hydrogen production.

Hydrogen production processes based on nonthermal plasma have the potential to provide significantly greater energy efficiency and/or selectivity than those based on thermal plasma. Plasma in industrial applications is typically generated by electrical discharges to bring a working gas (usually inert gases, nitrogen, or air) to a partially ionized state. In thermal plasma, the constitutive species, namely free electrons and so-called heavy-species (i.e., ions, excited and ground-state atoms and molecules), are in thermal equilibrium at a relatively high temperature, e.g., near 20000 K for arc discharge plasmas. In contrast, in nonthermal plasma, the free electrons are at significantly higher temperatures (typically between 1 and 10 eV, where 1 eV ~ 11600 K), than the heavy-species (from a few hundred to < 2000 K). The thermal nonequilibrium in nonthermal plasmas can translate into processes with higher energy efficiency and/or selectivity [20], [21], by directing the energy of electrons towards desired chemical reactions while limiting the energy carried by the gas species (which, although also driving chemical reactions, manifests as undesired heating). Additionally, nonthermal plasma processes are generally more amenable to compact, modular implementations, which may be favored for distributed (de-centralized) and small-scale installations.

The present study focuses on the use of a nonthermal plasma approach for the production of hydrogen from biomass and plastic waste. The model feedstocks for biomass and plastic waste are cellulose and low-density polyethylene (LDPE), respectively. Cellulose is a linear chain of repeated anhydroglucose rings $(C_6H_{10}O_5)_n$, usually between 10000 to 15000 units long [22], depending on the source material. The anhydroglucose units are bonded covalently by β-1,4' glycosidic links, which provide mechanical stiffness [23]. Cellulose is considered the most common organic compound on earth [24], naturally embedded in hemp, cotton, wood, crop residues, linen, etc. [22]. Investigations on the production of hydrogen from cellulose via electricity (rather than heat) have focused on electrochemical routes. Wei *et al*. [25] investigated electrochemically assisted molten carbonate pyrolysis of cellulose in the absence of a catalyst and obtained the maximum hydrogen yield of 8.3 mmol/g of cellulose at the relatively low temperature of 600 ºC. Similarly, Zeng *et al*. [26] studied molten salt pyrolysis of cellulose using a mixture of $Na_2CO_3$, $K_2CO_3$, and $Li_2CO_3$ as the electrolyte and achieved a peak hydrogen yield of 3.1 mmol/g of cellulose at 650 ºC. Furthermore, they also observed that hydrogen content increases rapidly from 18.05 to 26.19 vol.% as the result of increasing temperature from 650 to 850 ºC. In an experimental study of the transient behavior of devolatilization and char reactions during the steam gasification of biomass, Moon *et al*. [27] obtained a hydrogen production rate of about 800 mmol/h in ~ 2 minutes at an operating temperature of 700 ºC. Hoang *et al*. [28] characterized hydrogen production from steam gasification of plant-originated lignocellulosic biomass and obtained the maximum hydrogen yield of 55.6 mmol/g of cellulose at the operating temperature of 900 ºC.

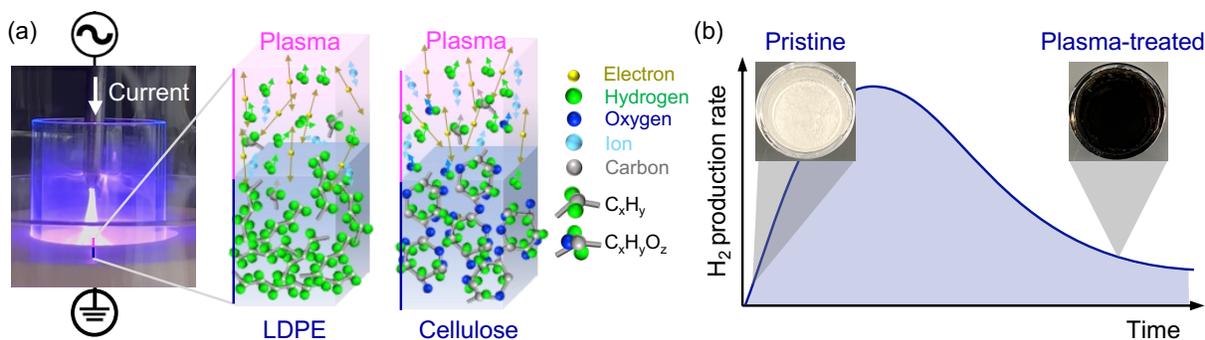

**Fig. 1. An overview of plasma dehydrogenation**. (a) Streamer Dielectric-Barrier Discharge (SDBD) plasma treatment of low-density polyethylene (LDPE) and cellulose, and (b) expected evolution of the hydrogen production rate as a function of treatment time.

Regarding the use of plastic waste for hydrogen production, Aminu *et al*. [29] reported a hydrogen production yield of 4.1 mmol/g from polyethylene during a two-stage low-temperature plasma catalytic treatment of plastic waste. Chai *et al*. [30] catalytically pyrolyzed a composite mixture of LDPE and pinewood dust using Ni-CaO-C as a catalyst and obtained an optimal hydrogen yield of 115.3 mmol/g of feedstock (LDPE to Pinewood dust ratio of 1 to 1) and hydrogen selectivity of 86.7% when 5 ml of water was injected, and the reactor was operating at 700 ºC. Nguyen and Carreon [31] investigated the catalytic deconstruction of high-density polyethylene (HDPE) via nonthermal plasma, reporting hydrogen yield of 8 mmol/g HDPE and selectivity of about 50%. In prior work by the authors [32], two nonthermal plasma reactors, based on transferred arc (transarc) and gliding arc (glidarc) discharges, were devised and used to produce hydrogen from LDPE. The maximum hydrogen yields were 0.33 and 0.42 mmol/g LDPE, with the corresponding minimum energy cost of 3100 and 3300 kWh/kg of $H_2$, for transarc and glidarc reactors, respectively. The study revealed that, despite the comparable yield and energy cost of these two largely different plasma sources, the transfer of electric current through the feedstock (as in the transarc reactor) leads to more compact and rapid-response processes.

In this study, the use of Streamer Dielectric-Barrier Discharge (SDBD) plasma, using argon as the working gas and operating at (near) atmospheric pressure, is investigated for the production of hydrogen from cellulose and LDPE as organic polymeric waste models. The approach is schematically summarized in Fig. 1. A high-voltage alternating-current (AC) power supply is used to generate plasma between a metal electrode and the feedstock placed over a dielectric barrier. SDBD plasma is a highly reactive medium composed of highly energetic electrons, ions, and metastable species. The charged species oscillate along the direction parallel to the electrode due to the imposed AC electric field. These reactive species interact with the cellulose and LDPE molecules, causing chain scissions and the release of hydrogen and low hydrocarbons as gas products, and the de-hydrogenation or carbonization of the remaining feedstock. Due to the nonthermal nature of SDBD plasma, the reactor operates at a relatively low temperature (< 200 ºC average inside the reactor chamber). The temperature is measured by an infrared thermometer with the laser beam incident on the surface of the feedstock. Given the decreasing availability of hydrogen to interact with plasma species as the process progresses, it is expected that the hydrogen production rate will initially increase until achieving a maximum and then monotonically decrease as the feedstock gets de-hydrogenated.

The article is organized as follows. Section 2 describes the experimental setup and procedures, as well as the electrical model and spectroscopic diagnostics of the SDBD reactor. In section 3, the results for hydrogen production from cellulose are presented, while section 4 consists of the results of hydrogen production from LDPE. Finally, the concluding remarks are presented in section 5.

## 2. Reactor design and experiment

### 2.1 Streamer Dielectric Barrier Discharge (SDBD) Reactor

The designed Streamer Dielectric Barrier Discharge (SDBD) reactor to produce hydrogen from polymeric solids is schematically depicted in Fig. 2. The reactor is designed with a pin-to-plate dielectric configuration and aimed to operate with plasma in a streamer (filamentary) discharge mode. The SDBD name is derived from the discharge mode of operation as well as the essential role of the dielectric barrier in the performance of the hydrogen production process. The reactor is powered by an alternating-current (AC) high-voltage power supply with a tungsten pin as the powered electrode (Fig. 2a). The voltage level of the power supply $V$ (from 0 to 100%) and the working gas flow rate $Q$ are the main operating parameters. The reactor's chamber has a diameter $D = 76$ mm and height of 160 mm (Fig. 2b). The pin electrode is electrically isolated by a ceramic bushing, and plasma is generated between the electrode's tip and the solid feedstock, which is placed in a crucible assembly on top of an aluminum plate acting as the ground electrode. The crucible consists of two quartz dishes - the larger dish has a diameter of 56 mm, while the smaller one, which contains the feedstock, has a diameter $d = 20$ mm and includes an annular dielectric ring made of alumina. This ensures sufficient electrical insulation to mitigate undesired arcing (i.e., the formation of an electrical discharge circumventing the feedstock). The geometrical dimensions of the crucible assembly are of primary importance in determining the characteristics of the plasma and the performance of the process. Particularly, in the absence of the dielectric barrier, the discharge is weaker, and so is the rate of hydrogen production. The presence of the dielectric increases the amount of electrical power deposited on the feedstock. The main dimensions of the plasma-feedstock-dielectric barrier assembly are shown in Fig. 2c, namely electrode-feedstock spacing $h_p$ = 5 mm, feedstock height $h_f$ (different size of cellulose and LDPE, see section 2.2), and dielectric height $h_d$ = 4.5 mm.

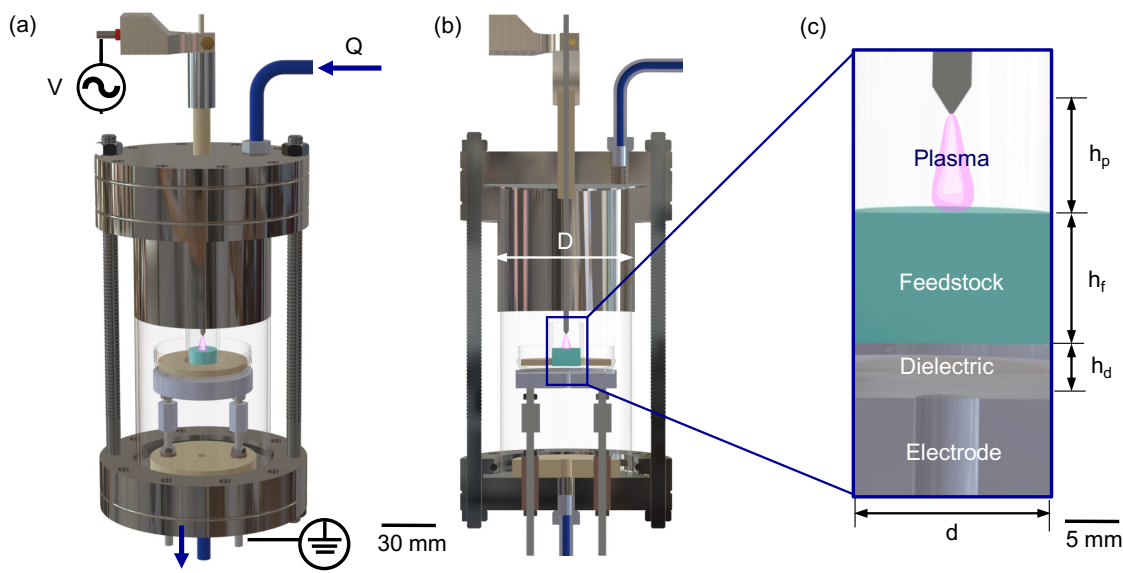

**Fig. 2. SDBD plasma reactor design based on pin-to-plate streamer discharge – dielectric barrier configuration**. (a) Reactor assembly with voltage level from the power supply $V$ and processing gas flow rate $Q$ as the main process parameters. (b) Cross-sectional view of the reactor depicting the reactor chamber's diameter $D$. (c) Zoom-in view of the plasma-feedstock-dielectric region, depicting the electrode-feedstock spacing $h_p$, feedstock height $h_f$, dielectric height $h_d$, and diameter of the feedstock holder $d$.

## 2.2. Experimental setup and procedures

*Components.* The SDBD reactor generates streamer plasma in contact with the solid feedstock, leading to the production of hydrogen and carbon co-products. The experimental setup is depicted in Fig. 3. The reactor is operated by a high-voltage AC power supply (PVM 500-2500) characterized by an adjustable peak-to-peak voltage of 1-40 kV and a maximum current of 25 mA. Argon is used as the processing gas, and two mass flow controllers are used to measure and control the inlet and outlet flow rates. The electrical characteristics of the reactor's electrical circuit are measured using a Tektronix oscilloscope (TBS 2104) connected to a current probe (P6021A) and to a high-voltage probe (P6015A). To measure the spectral characteristics (optical emission) of the plasma, an Avantes spectrometer (ULS-2048-USB2) equipped with a set of optical lenses and a fiber optic probe is used. The gas products are collected in sampling bags and subsequently analyzed by a Shimadzu-2014 gas chromatograph. The morphological structure of the solid samples before and after plasma treatment is characterized using a digital camera and a field emission scanning electron microscope (FESEM) JSM 7401 coupled with an energy dispersive x-ray spectroscopy (EDS) detector. The elemental composition of the solids sample is determined by the CHN-elemental analysis.

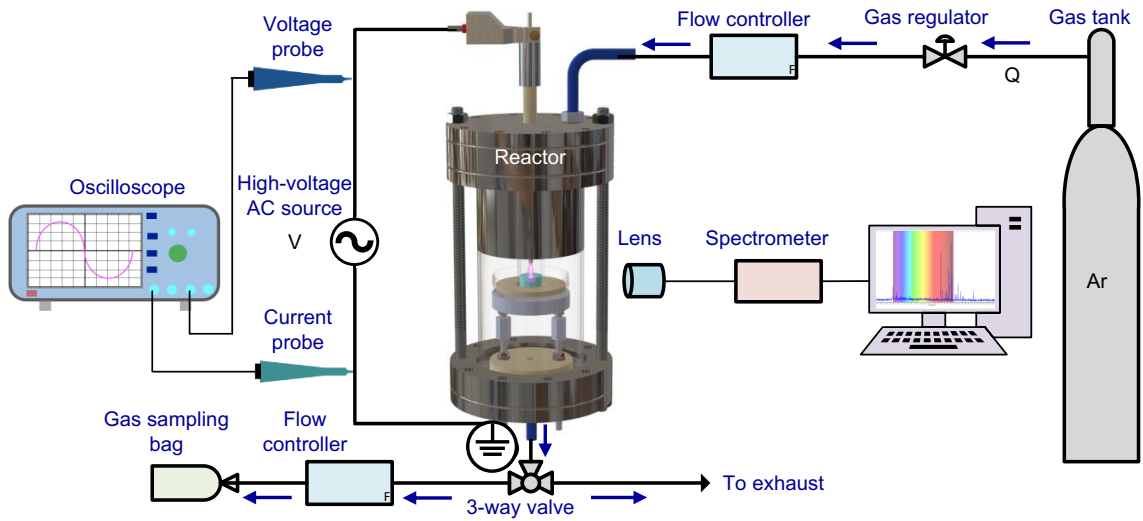

**Fig.3 Experimental setup for hydrogen production from cellulose and LDPE via SDBD plasma.** Schematic of the experimental layout showing the SDBD reactor, the different diagnostics, and the gas and electrical lines.

*Feedstock sample preparation.* The cellulose feedstock is a powder made of cotton linters commercially available from Millipore Sigma (Supelco, V001141). The sample is prepared by mechanically pressing 1 g of cellulose in the inner crucible plate (diameter $d = 20$ mm and height $h_f \sim 10$ mm, see Fig. 2c) using a manual hydraulic pellet press. The LDPE feedstock consists of 1 g of pellets with an average diameter of 3 mm from Millipore Sigma (Sigma Aldrich, 428 043), which are pre-melted at 180 ºC for 15 min and then re-solidified in the inner crucible (diameter $d = 20$ mm and height $h_f \sim 6$ mm, see Fig. 2c).

*Experimental procedure.* In each experiment, cellulose and LDPE are treated with SDBD plasma for varying treatment times with the same voltage level $V = 60\%$ and inlet flow rate $Q = 0.01$ slpm of argon. The 60% voltage level was chosen as representative of the process leading to root-mean-square input power, $P_{t,rms} = 50$ W and 53 W for cellulose and LDPE, respectively. Given the faster carbonization of cellulose, the treatment times of the experiments are set as $t_{treatment} = \{\ 0.5, 1, 2, 3, 4, 5, 6, 15\ \}$ min. For LDPE, the treatment time is more uniformly spread compared to cellulose and set as $t_{treatment} = \{\ 0.5, 1, 2, 4, 6, 8, 10, 15\ \}$ min. The electrical characteristics are measured at different time instants $t_{elec}$ depending on the treatment time. Specifically, for cellulose, for $t_{treatment} = 0.5$ min, $t_{elec} = \{\ 0.2\ \}$ min; for $t_{treatment} = 1$ min, $t_{elec} = \{\ 0.2, 0.4\ \}$ min; and for $t_{treatment} = \{\ 2, 3, 4, 5, 15\ \}$ min, $t_{elec} = \{\ 0.5, 1\ \}$ min. For LDPE, for $t_{treatment} = 0.5$ min, $t_{elec} = \{\ 0.2\ \}$ min; for $t_{treatment} = 1$ min, $t_{elec} = \{\ 0.2, 0.4\ \}$ min; and for $t_{treatment} = \{\ 2, 4, 6, 10, 15\ \}$ min, $t_{elec} = \{\ 0.5, 1\ \}$ min. The same time instants are considered for measuring the reactor's temperatures using an infrared thermometer. All the generated gas products are collected throughout the plasma treatment (time $t$ from 0 to $t_{treatment}$) with the constant flow rate $Q = 0.01$ slpm, and after treatment ($t > t_{treatment}$) with a higher flow rate of $Q = 0.1$ slpm for 15 min for purging. The purging with argon drives all the generated gaseous products into the sampling bag for analysis by gas chromatography (GC). The spectroscopic measurements are carried out 15 seconds from the beginning of each experiment before the emanation of the opaque gaseous products that obscure the optical access of the plasma (see section 3 and section 4).

***Gas product analysis.*** The identification and quantification of the gaseous products are performed using a Shimadzu GC-2010 Plus GC system equipped with a mass spectrometer (MS) and a flame ionization detector (FID) and a Shimadzu GC-2014 system equipped with a thermal conductivity detector (TCD), and FID. For hydrocarbon products analysis, 1 mL of the sample is injected into a Restek packed column (80486-810, 2 m) connected to a TCD, and another 1 mL injected that is connected to the FID. Hydrogen is used as the carrier gas through the column, with a flow rate set at 12.50 ml/min. Peak identification and calibration are achieved from results obtained using standard gas mixtures containing CO, $CO_2$, and $C_1-C_4$ hydrocarbons (Scotty Specialty Gases). The GC is programmed with an injection temperature of 250 °C, and the programmed temperature regime for the GC oven is as follows: start at 35 °C ramp up to 50 °C at 7.5 °C/min, hold for 1 min, ramp up to 100 °C at 5 °C/min, hold for 3 min, ramp up to 200 °C at 10 °C/min, hold for 7 min, ramp up to 250 °C at 25 °C/min, and hold at 250 °C for 10 min. For hydrogen products, 2 mL of the sample is injected into a Restek packed column (80486-810, 2 m) connected to a TCD. Argon is used as the carrier gas through the column, with a flow rate set at 12.50 ml/min. The GC is programmed with an injection temperature of 250 °C, and the programmed temperature regime for the GC oven is: start at 35 °C ramp up to 75 °C at 7.5 °C/min.

## 2.3. Electrical model

An electrical model allows the determination of the power consumed by the plasma given the total power consumed measured by the oscilloscope. The model assumes that the plasma within the discharge gap (electrode-feedstock spacing $h_p$), feedstock, and dielectric, and can be electrically described as parallel-plate capacitors in series, as schematically depicted in Fig. 4a. Based on this assumption, the electrical capacitances of the components in the plasma circuit are defined by their geometrical configurations as

$$C_f = \frac{A_f}{h_f} \kappa_f \varepsilon_0 \tag{1}$$

and

$$C_d = \frac{A_d}{h_d} \kappa_d \varepsilon_0, \tag{2}$$

where $C_f$ and $C_d$ are the capacitances of the feedstock and dielectric, respectively; $A_f$ and $A_d$ are the cross-sectional areas of feedstock and dielectric, respectively; $\varepsilon_0$ is the permittivity of free space; and $\kappa_f$ and $\kappa_d$ are the dielectric constants of feedstock and dielectric, respectively. The dielectric constant for the dielectric crucible made of quartz is $\kappa_d = 3.8$ [33], whereas the dielectric constant of the feedstock is $\kappa_f = 2.5$-2.6 [34] for cellulose and $\kappa_f = 2.2$-2.35 [35] for LDPE.

Using Kirchhoff's law, the total input voltage $U_a(t)$ expressed in terms of plasma voltage $U_p(t)$, voltage across the dielectric $U_d(t)$, feedstock voltage $U_f(t)$, and effective dielectric voltage $U_D(t)$ is given by:

$$U_a(t) = U_p(t) + U_f(t) + U_d(t) = U_p(t) + U_D(t). \tag{3}$$

The total input current $I_t(t)$ is the sum of plasma current $I_p(t)$ and the displacement current through the gap $I_{p,g}(t)$, i.e.,

$$I_t(t) = I_p(t) + I_{p,g}(t). \tag{4}$$

The effective capacitance of the feedstock and dielectric $C_D$, given that these are assumed to operate in series, is given by:

$$\frac{1}{C_D} = \frac{1}{C_f} + \frac{1}{C_d}, \tag{5}$$

where $C_f$ and $C_d$ are the capacitances of the feedstock and dielectric, respectively. Liu and Neiger [36] derived the voltage across the feedstock and dielectric, as:

$$U_D(t) = \frac{1}{C_D}\int_0^t I_{t,a}(t') \, dt' + U_D(0), \tag{6}$$

where $U_D(0)$ is the memory voltage, which depends on an arbitrarily zero set time ($t = 0$) and is attributed to the memory charges deposited during the preceding AC voltage cycle. Considering that the negative voltage peak occurs at time zero, $U_D(0)$ becomes a constant and is defined in terms of the period $T$, i.e.,

$$U_D(0) = -\frac{1}{2C_D}\int_0^{\frac{T}{2}} I_{t,a}(t') \, dt'. \tag{7}$$

The plasma discharge current $I_p(t)$ can be determined from the input current by

$$I_p(t) = \left(1 + \frac{C_{p,g}}{C_D}\right) I_{t,a}(t) - C_{p,g}\frac{dU_a(t)}{dt}, \tag{8}$$

where the first and second terms on the right-hand side represent the total displacement current $I_{v,g}(t)$ and the gap displacement current $I_{p,g}(t)$, respectively. The total displacement current, sometimes referred as the external discharge current, is attributed to the effective capacitance of the plasma-gap, feedstock, and dielectric. Hence, it is generally erroneous to assume that the input current is the same as the plasma current, even when the gap displacement current is small and can be neglected.

The instantaneous input power can be calculated as

$$P_t(t) = U_a(t) I_t(t), \tag{9}$$

whereas the instantaneous plasma power is determined by

$$P_p(t) = U_p(t) I_p(t). \tag{10}$$

In the SDBD reactor, describing the plasma as mostly acting capacitively, with plasma-gap capacitance $C_{p,g}$, is a substantial approximation. Nevertheless, such an approximation is consistent with more conventional DBD electrical models and can be considered as reasonable as a first-order approximation to estimate the power consumed by the plasma. Adopting this approximation, $C_{p,g}$ is approximated as a constant determined from the experimentally-measured electrical response of the system by considering a fixed input power and calculating the range for which the plasma power is less or equal to the input power. The approach is schematically shown in Fig. 4b. The plasma power decreases to a minimum and subsequently increases monotonically with increasing $C_{p,g}$. This is attributed to the variation of gap displacement current $I_{p,g}(t)$ and total displacement current $I_{v,g}(t)$, hence changing the plasma current $I_p(t)$ and, subsequently, the plasma power $P_p(t)$. When $C_{p,g}$ is zero, the gap displacement current is zero, and hence the plasma power equals the input power. As $C_{p,g}$ increases, $I_{p,g}(t)$ increases more rapidly than $I_{v,g}(t)$, leading to a reduction in $I_p(t)$, and consequently to a decrease in plasma power $P_p(t)$. Further increasing $C_{p,g}$ leads to greater $I_{v,g}(t)$ than $I_{p,g}(t)$, and as a result, $P_p(t)$ also increases. The minimum point suggests that the increase in $I_{v,g}(t)$ and gap displacement current are equal.

Using experimental $U_a(t)$, $I_t(t)$, and $P_t(t)$ data during the treatment of cellulose and given that the plasma power cannot exceed the total power, acceptable values of $C_{p,g}$ are found within the range $0.1 \leq C_{p,g} \leq 2.9$ pF. This range of capacitance across the gap leads to corresponding plasma power between in the range $90\%P_{t,rms} < P_{p,rms} < 100\%P_{t,rms}$, where $P_{p,rms}$ is the root-mean-square (rms) of plasma power. The estimated range is comparable with the results by Ozkan *et al*. [37], which obtained 92% of absorbed power using the Lissajous method. Valdivia-Barrientos *et al*. [38] reported that the plasma voltage is 98% of the applied voltage, consistent with our model estimates. Therefore, during the operation of the SDBD reactor, it is expected that between 90 and 100% of the total power is consumed by the plasma. The oscillograms presented in Fig. 4c depicting the input voltage, input current, and input power of the SDBD plasma are measured by the oscilloscope. The SDBD plasma produces a smooth sinusoidal voltage signal that is out of phase with current and power. The sharply varying current has a ripple effect on the input power, leading to the spiky current and power waveforms.

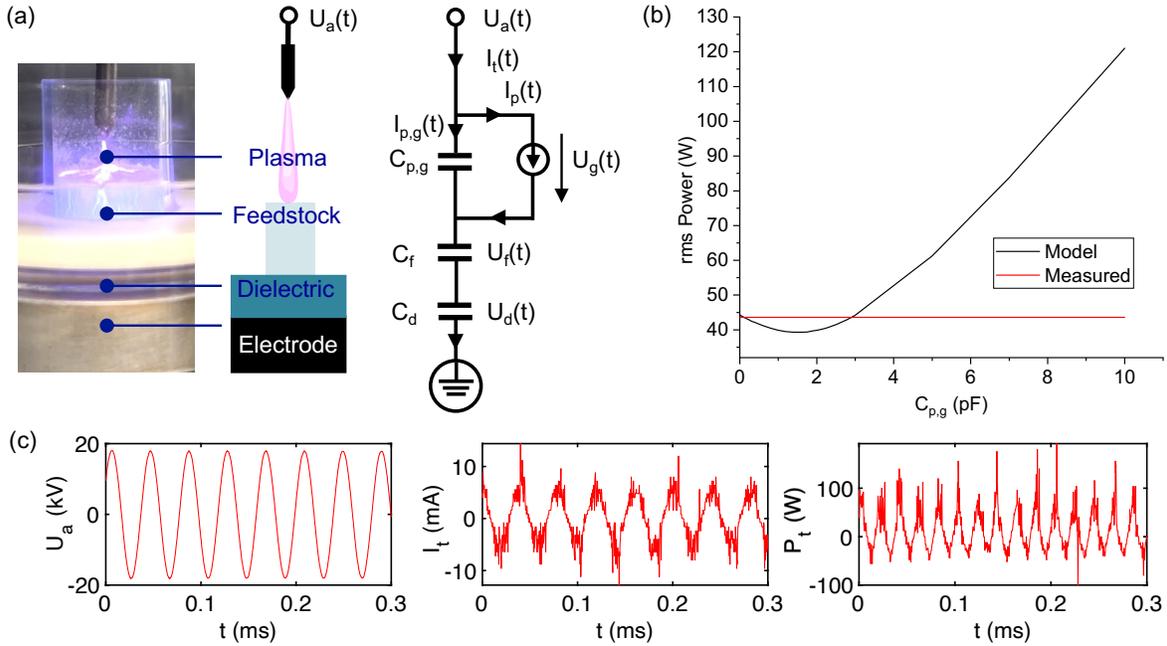

**Fig. 4. SDBD electrical model.** (a) Operation of the reactor during cellulose treatment and equivalent circuit diagram, with the feedstock and main reactor components represented as parallel plate capacitors. (b) Measured input power and the model as functions of capacitance plasma-gap capacitance $C_{p,g}$. (c) Oscillograms of input electrical characteristics of voltage, current, and power.

## 2.4. Spectroscopic diagnostics

Optical Emission Spectroscopy (OES) is used to determine the primary characteristics of the SDBD plasma, namely, representative values of excitation temperature $T_{exc}$, electron temperature $T_e$, and electron number density $n_e$. A spectrometer (Avantes ULS2048-USB2) with a wavelength range of 200-1100 nm and a grating of 300 lines/mm is used to measure the spectral emission from the plasma. The measurements are performed near the beginning ($t = 15$ s) of the feedstock treatment. A representative spectrum is shown in Fig. 5a obtained during the treatment of cellulose feedstock under operational conditions of voltage level $V = 60\%$ ($P_{t,rms} = 50$ W), flow rate $Q = 0.01$ slpm, and electrode-feedstock spacing $h_p = 5$ mm. The peaks represent the relative intensity of radiative transitions, from some upper energy level $i$ to a lower energy level $j$, with the wavelengths of primary transitions used for the analysis indicated within the figure.

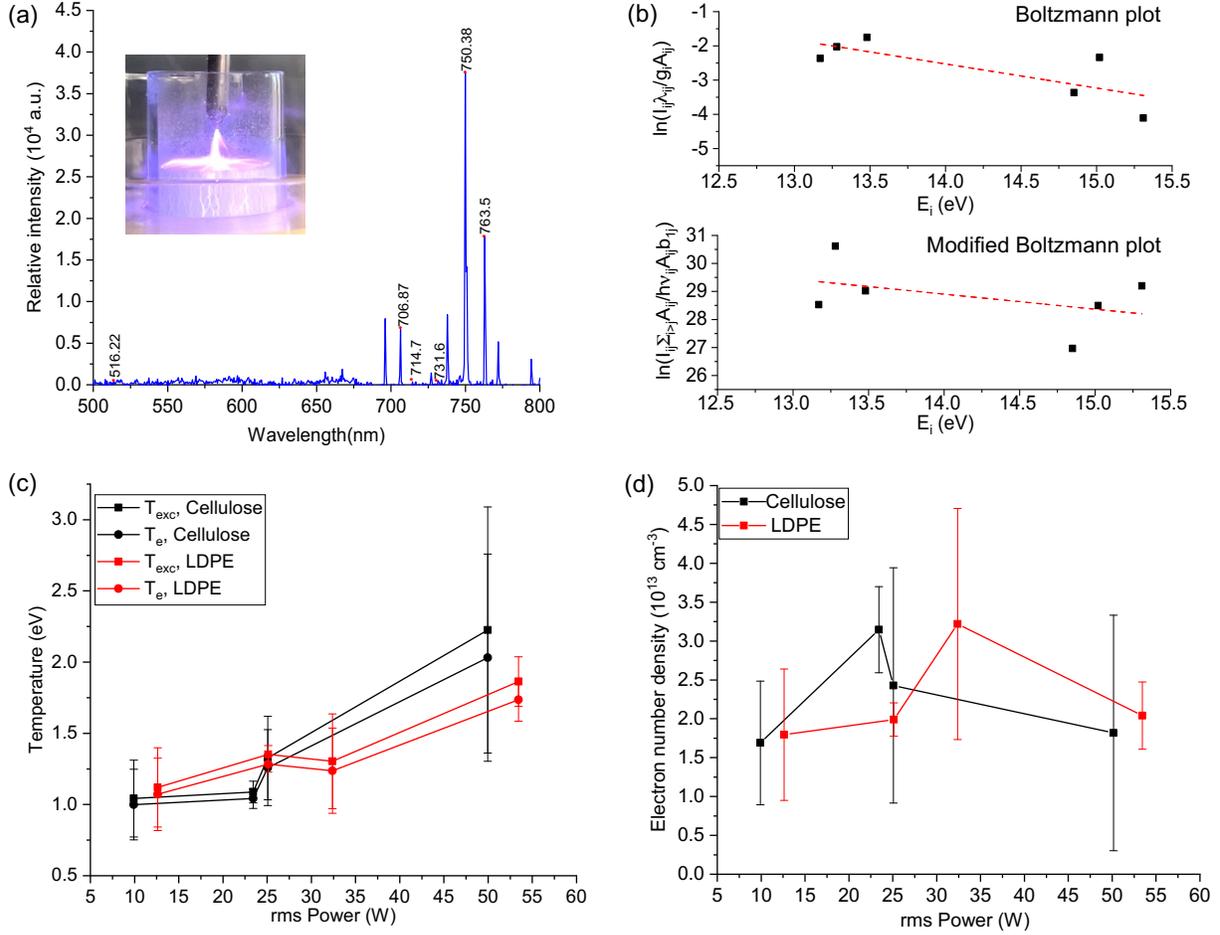

**Fig. 5. Spectroscopic characterization of SDBD plasma.** (a) Plasma spectra at the beginning of treatment of cellulose ($t$ = 0.5 min) and representative Ar I spectral lines and corresponding (b) Boltzmann plot for the determination of the excitation temperature ($T_{exc}$ = 1.42 eV) and modified Boltzmann plot for calculation of electron temperature ($T_e$ = 1.35 eV). (c) Excitation and electron temperatures and (d) electron number density $n_e$ as a function of power during the treatment of cellulose and LDPE.

The estimation of plasma properties from OES data in the present work is based on the use of the Boltzmann plot method, which has been extensively used for determining $T_e$ and $T_{exc}$ in a wide range of plasmas [39]–[42]. The method is based on comparing the relative intensity of representative thermometric species. The Boltzmann plot method assumes local thermal equilibrium in which the excitation and de-excitation mechanism is controlled by the electronic collisions, and both $T_e$ and $T_{exc}$ are the same. The excitation temperature $T_{exc}$ is calculated from the slope of the best fit of a Boltzmann distribution of excited states with the upper energy level $i$ and lower energy level $j$, leading to the expression [39]:

$$ln\left(\frac{I_{ij}\lambda_{ij}}{g_i A_{ij}}\right) = -\frac{E_i}{k_B T_{exc}} + D, \qquad (11)$$

where $g_i$ is the statistical weights of the upper level $i$ of the transition considered, $A_{ij}$ is the transition probability of the emitted spectra, $I_{ij}$ is the relative intensity of the spectral emission for upper to lower states, $\lambda_{ij}$ is the wavelength of the emitted spectra, $E_i$ is the excitation energy, $k_B$ is the Boltzmann constant, $T_{exc}$ is excitation temperature in electron volt (eV), and $D$ is a data-fitting constant.

Since the SDBD plasma is in a state of thermal nonequilibrium, i.e., different electron temperature from the heavy-species temperature – an intrinsic characteristic of nonthermal plasma, electronic collisions might not be the only processes controlling the excitation and de-excitation mechanism [39]. To estimate the electron temperature $T_e$, the modified Boltzmann plot method developed by Gordillo *et al.* [39] and used by several authors [40], [42] is adopted. This approach assumes that the plasma is in the state of corona balance in which the populating and depopulating mechanisms are attributed to electron-impact collisional excitation from the ground state and spontaneous radiative emission, respectively, and that the two mechanisms are balanced. Gordillo *et al.* [39] modified Boltzmann plot method leads to the equation:

$$ln\left(\frac{I_{ij}\sum_{i>j}A_{ij}}{h_P v_{ij}A_{ij}b_{1i}}\right) = -\frac{E_i}{k_B T_e} + B, \qquad (12)$$

where $h_P v_{ij}$ is the energy gap between levels $i$ and $j$ (with $h_P$ as the Planck constant and $v_{ij}$ the collision frequency between species), $b_{1i}$ is a constant function of the electron-impact excitation rate coefficient, $\sum_{i>j}A_{ij}$ is the summation of the transition probabilities starting from the upper energy level $i$, and $B$ is a data-fitting constant. The quantity $\sum_{i>j}A_{ij}$ is determined by considering all the possible spontaneous radiative transitions from the upper energy levels associated with the measured lines and then summing up their respective transition probabilities [39]. The parameters $I_{ij}$, $A_{ij}$, and $E_i$ are obtained from NIST [43] atomic spectra database.

The excitation temperature $T_{exc}$ and the electron temperature $T_e$ are determined from the slope of the least-squares fit of equation (11) and equation (12), respectively, for a given number of spectral lines (radiative transitions). In this study, six Ar-I lines corresponding to wavelengths: 516.2 nm (electronic transition 6d→4p), 706.9 nm (6s→4p), 714.7 nm (4p→4s), 731.6 nm (6p→4s), 750.4 nm (4p→4s), and 763.5 nm (4p→4s) are chosen, as indicated in Fig. 5a. The lines are chosen such that they have the greatest gap between the upper energy levels (levels $i$) at the expense of considering higher relative intensities. This reduces the error in the estimation of electron excitation and electron temperature resulting from smaller differences (of 1 eV or less) between the upper energy levels of the transitions [39]. The Boltzmann plot and modified Boltzmann plot for the determination of $T_{exc}$ and $T_e$, respectively, are presented in Fig 5b for operating conditions: voltage level $V$ = 60% ($P_{t,rms}$ = 50 W), $Q$ = 0.01 slpm, and $h_p$ = 5 mm. The least-squares linear fitting is applied to obtain the slopes for each plot, leading to $T_{exc}$ = 1.42 eV and $T_e$ = 1.35 eV. Three different spectral measurements are performed, and the average $T_{exc}$ and $T_e$, as well as the error, are determined and presented in Fig. 5c.

The dependence of $T_{exc}$ and $T_e$ with $P_{t,rms}$ during the treatment of cellulose and LDPE are shown in Fig. 5c. Both $T_{exc}$ and $T_e$ increase monotonically with $P_{t,rms}$ as the result of increased electron energy [42]. As $P_{t,rms}$ increases, the electrons gain more energy leading to the generation of a greater number of active species through inelastic electron-molecular species collisions [44]. The small difference in $T_{exc}$ and $T_e$ observed in Fig. 5c is consistent with reports by other authors [39], [40], [42]. The results in Fig. 5c show that $T_{exc}$ and $T_e$ in the treatment of both cellulose and LDPE are similar, implying comparable plasma conditions and that the type of feedstock has a relatively minor role in the plasma characteristics of the system.

The electron number density $n_e$ of the SDBD plasma is estimated using the approach by Kais *et al.* [40]. Their approach provides an expression for $n_e$ relating the sheath potential $V_{sh}$, the ionization energy $E_{ion}$ of the gas (15.7 eV in the case of argon), and the electron temperature according to:

$$n_e = \frac{P_{t,rms}}{A_s}\left(\left(\frac{k_BT_e}{2\pi m_e}\right)^{\frac{1}{2}}\exp\left(\frac{eV_{sh}}{k_BT_e}\right)(2k_BT_e+E_{ion})+0.3k_BT_e\left(\frac{k_BT_e}{m_i}\right)^{\frac{1}{2}}\left(\frac{k_BT_e}{2}\right)\left|\ln\left(\frac{2\pi m_e}{m_i}\right)+1\right|\right)^{-1}, \quad (13)$$

where $A_s$ is the substrate cross-sectional area, $e$ the elementary charge, $m_e$ electron mass, and $m_i$ the ion mass. Consistent with derivation leading to equation (13), the sheath potential $V_{sh}$ can be determined using the expression [40]:

$$V_{sh} = \left(\frac{k_BT_e}{2e}\right)\ln\left(\frac{m_i}{2\pi m_e}\right). \quad (14)$$

The electron number density for an electron temperature of 1.35 eV and $P_{t,rms}$ = 50 W of the SDBD plasma used in the production of hydrogen from cellulose is $1.82\times10^{13}$ cm$^{-3}$. The dependence of the average electron number density $n_e$ as a function of power $P_{t,rms}$ during the treatment of cellulose and LDPE is shown in Fig. 5d. The results show that $n_e$ remains approximately constant with varying power, with the lowest value of $1.69\times10^{13}$ cm$^{-3}$ corresponding to $P_{t,rms}$ = 10 W during the treatment of cellulose. The maximum $n_e$ of $3.33\times10^{13}$ cm$^{-3}$ is attained during the treatment of LDPE with $P_{t,rms}$ = 32 W. The estimated $n_e$ range is comparable with that reported for other nonthermal plasma, typically in the order of $10^9$ to $10^{13}$ cm$^{-3}$ [39]–[41], [45].

## 3. Results and discussion

### 3.1. Hydrogen production from Cellulose

***SDBD plasma-feedstock interaction.*** The interaction between SDBD plasma and cellulose at the end of each treatment is depicted in Fig. 6. The treatment process initially leads to the emission a mainly purple glow characteristic of argon plasma (as shown in Fig. 5a). As the plasma treatment progresses, the purple glow transitions to yellow for $t$ = 0.5 min. Moon *et al*. [27], [46] noted that devolatilization and char reaction are the main regimes during biomass gasification. Therefore, the glow transition observed for $t$ = 0.5 min probably indicates the beginning of carbonization. The rapid devolatilization at the beginning of the process is probably due to concerted reactions involving glycosidic oxygen atoms with an activation energy of about 210-230 kJ/mol [47]. This results in glycosidic bond cleavage in any random positions in the cellulose chain leading to the formation of anhydrosugars which are further converted into char and gaseous products such as $CO_2$, $H_2O$, $H_2$, and CO [48].

The images in Fig. 6 also show the production of fine particles depicted by the clouding of the reactor chamber (smoke), particularly noticeable at $t$ = 1 min. The intensity of the smoke and yellow emissions increases with time for $t$ = 1 to 4 min, then it decreases. The increase in smoke and yellow emissions is probably attributed to the dominance of the devolatilization of the feedstock and the presence of oxygen atoms, respectively. However, the shift of devolatilization to char reaction could be responsible for the decreased smoke intensity. The red glow observed after $t$ = 5 min probably depicts the emission from carbon particles.

Although the precise mechanism for the carbonization of cellulose via pyrolysis is not fully understood, previous experimental investigations provide important insights [49]. It is suggested that the initial step of the carbonization of cellulose is the devolatilization of cellulose into char residues and volatiles, i.e., $(C_6H_{10}O_5)_n$ + Plasma → Char + Volatiles [50]–[52]. The reaction mechanism involves the cleavage of the C-O glycosidic bond, which then proceeded by multiple homogenous reactions leading to the formation of anhydrosugars which are converted into char and gaseous products such as $CO_2$, $H_2O$, $H_2$, and CO [48].

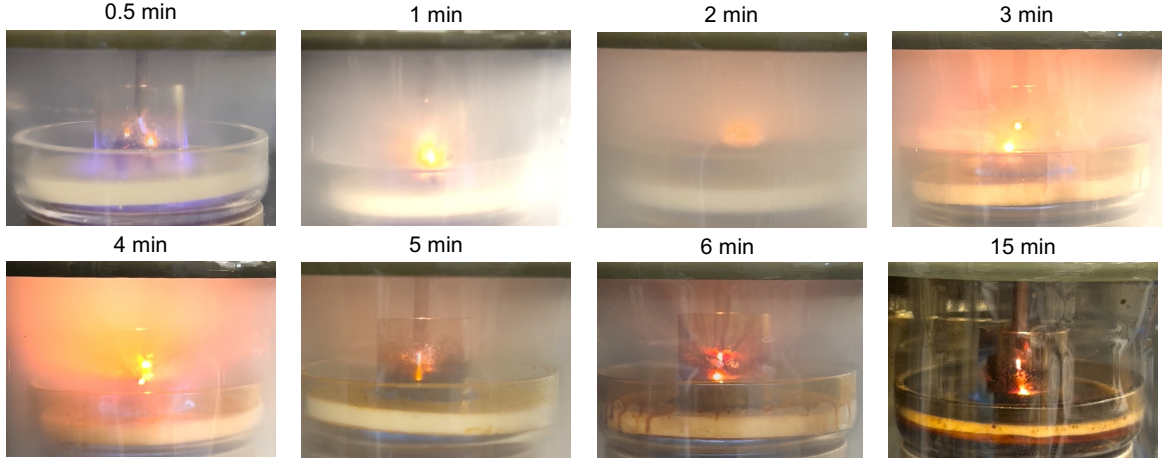

**Fig. 6. Optical imaging of plasma treatment of cellulose.** SDBD plasma interaction with cellulose at different times under operation conditions of $V = 60\%$ ($P_{t,rms} = 50$ W) and $Q = 0.01$ slpm.

*Hydrogen production and production efficiency.* The performance of the SDBD reactor during the production of hydrogen from cellulose (cellulose dehydrogenation or carbonization) is next assessed. The set-up (Fig. 3) is operated at near atmospheric pressure. The temperature of the reactor chamber is measured using an infrared thermometer, ranging from 21 to 66 °C, depending on the time with a treatment (between 0.5 and 15 min), confirming the relatively low temperature of the process.

The reactor's performance parameters considered are *Cumulative $H_2$ production*, *Mean $H_2$ production rate*, *$H_2$ production efficiency*, and *Energy cost of $H_2$ production*. The *Cumulative $H_2$ production* quantifies the total amount of hydrogen collected in the sampling bag and analyzed via GC, and *Mean $H_2$ production rate = Cumulative $H_2$ production/$t_{treatment}$*. *$H_2$ production efficiency* (in units of mol/kWh) is defined as

$$H_2 \text{ Production efficiency} = \frac{\text{Mean } H_2 \text{ Production rate}}{\text{Input rms power}} \quad (15)$$

and the *Energy cost of $H_2$ production* (kWh/kg $H_2$) is determined from

$$\text{Energy cost of } H_2 \text{ production} = \frac{1000}{M_w(H_2 \text{ Production efficiency})}, \quad (16)$$

where $M_w$ is the molecular weight of hydrogen and 1000 is the conversion factor for grams to kilograms.

The obtained performance of the SDBD reactor as a function of treatment time is depicted in Fig. 7. The *Cumulative $H_2$ production*, namely the total amount of hydrogen produced for each $t_{treatment}$, is presented in Fig. 7a. The *Cumulative $H_2$ production* rapidly increases for $t_{treatment} < 3$ min of and then depicts a significantly slower increase. As observed by Sun *et al*. [46] and indicated in section 3.1, the gas release process from biomass gasification occurs mainly in two regimes: devolatilization and char reaction. Devolatilization is the main process of gas release in cellulose gasification since cellulose is composed of largely volatile components. This can explain the sharp rise in cumulative hydrogen production in the first 3 minutes of treatment, which is comparable to the behavior observed by Moon *et al*. [27], who obtained a peak devolatilization time of 4 minutes for biomass steam gasification with an operating temperature of 900 °C. The char reaction is a slow process but lasts for a longer time [46], which can explain why the cumulative hydrogen production remains almost constant during the rest of the treatment time. Overall, the trend of cumulative hydrogen production is comparable to hydrogen production from biomass pyrolysis at 900 °C obtained by Moon *et al*. [27].

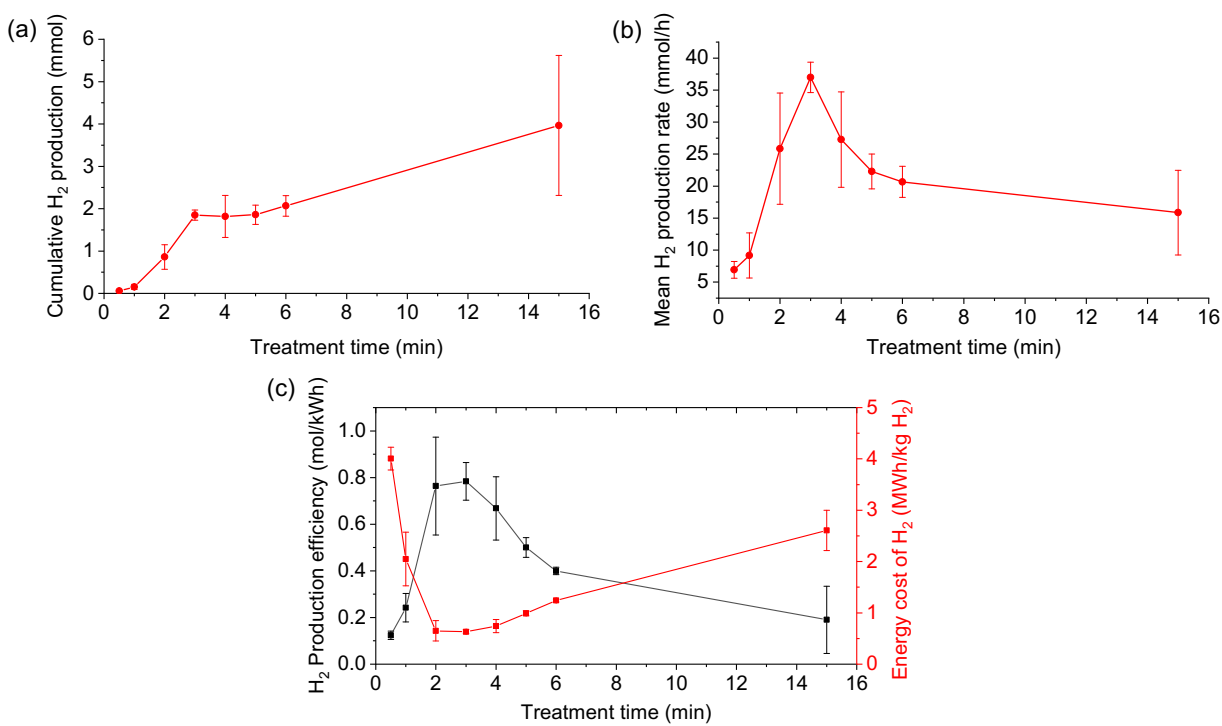

**Fig. 7. Hydrogen production performance of the SDBD plasma treatment of cellulose.** (a) Cumulative hydrogen production, (b) mean hydrogen production rate, (c) hydrogen production efficiency, and energy cost of hydrogen production versus treatment time.

As hypothesized in section 2, the mean hydrogen production rate increases with treatment time to a maximum, reaching 40 mmol/h in 3 minutes of treatment before decreasing and eventually remaining almost constant, as illustrated in Fig. 7b. The rapid increase in mean hydrogen production rate is likely attributed to the reactions involving glycosidic oxygen atoms leading to the formation of anhydrosugars, mainly levoglucosan, which is further converted to hydrogen and other gaseous products, and char [48]. Moon *et al*. [27] observed similar behavior during the transient production of hydrogen from biomass via pyrolysis at 600 °C with peak hydrogen production rate of about 800 mmol/h. The occurrence of the peak production suggests that the plasma treatment of cellulose leading to syngas production is a two-stage process of devolatilization and char reaction, as noted by other authors [27], [46], [53]. The devolatilization, which accounts for the main process of gas release, dominates during the first 3 minutes of treatment. However, the char reaction becomes more significant as the cellulose treatment progresses, leading to decreasing mean hydrogen production rate. The peak mean hydrogen production rate at 3 minutes seems to indicate the maximized synergistic effect of both devolatilization and char reaction, as noted in [27].

Hydrogen production efficiency quantifies the amount of energy required to produce a unit quantity of hydrogen. As shown in Fig. 3c (black line), the *$H_2$ Production efficiency* increases rapidly to a maximum of 0.8 mol/kWh in 3 minutes and then starts decreasing. This efficiency is an order of magnitude smaller than the 2.1 mol/kWh obtained by Wu *et al*. [54] in the DBD plasma reforming of n-pentane. Similarly, Panda and Sahu [55] obtained an efficiency of 1.2 mol/kWh from methanol using DBD plasma with conducting sea water as an electrode. The occurrence of maximum efficiency is also a manifestation of the existence of the two regimes of devolatilization and char reaction of biomass pyrolysis, as reported in [53], in which devolatilization dominates the initial stage of the process and char reaction becomes more pronounced after the peak efficiency is reached. The peak production efficiency, similar to the peak hydrogen production, is attributed to the synergistic effect of devolatilization and char reaction.

The *Energy cost of $H_2$ production*, defined as the amount of electrical energy required to produce 1 kg of hydrogen as shown in Fig. 3c (red line). The *Energy cost of $H_2$ production* decreases rapidly to a minimum of 630 kWh/kg $H_2$ for $t_{treatment}$ = 3, consistent with the maximum *$H_2$ Production efficiency*, as well as the maximum *Mean $H_2$ production rate*. Comparably, the energy cost of hydrogen production is about two orders of magnitude greater than for water electrolysis (47.6 kWh/kg) [56] and methane steam reforming (21.9 kWh/kg) [57]. The higher energy cost of producing hydrogen from cellulose is expected, given the greater embedding of hydrogen within a polymeric solid. Nevertheless, the plasma valorization of solid feedstock such as cellulose, representative of biomass waste, could lead to environmental and economic benefits.

***Gas product yield and hydrogen selectivity.*** To assess the selectivity of the production of hydrogen from cellulose via SDBD plasma, the area of the main gas products detected by gas chromatography is analyzed for the representative treatment times of $t_{treatment}$ = 3 min and $t_{treatment}$ = 15 min. The main gas products obtained during the plasma treatment of cellulose are $H_2$, $CO$, $CO_2$, $CH_4$, $C_2H_4$, and $C_2H_6$. The 3-minute treatment depicts the greatest hydrogen production efficiency and minimum energy cost (Fig. 7), whereas the 15-minute treatment depicts the greater charring (carbonization) of the feedstock.

*Gas product yield* is defined as the amount of gas product produced per gram of cellulose during the plasma treatment, and it is quantified by

$$Gas\ product\ yield = \frac{Moles\ of\ gas\ product}{Total\ mass\ of\ feedstock}. \tag{17}$$

Similarly, *Selectivity* is derived from *Gas product yield* as

$$Selectivity = \frac{Gas\ product\ yield}{Total\ gas\ product\ yield} \times 100\%. \tag{18}$$

The *Gas product yield* for the plasma treatment of cellulose under the two treatment times of 3 and 15 min is presented in Fig. 8a. The hydrogen yield for $t_{treatment}$ = 3 min and 15 min is 1.8 and 4.0 mmol/g of cellulose, respectively, and it is significantly higher than the yield of other gas products. These results are comparable to the electrolytic pyrolysis of biomass. For instance, Zeng *et al.* [26] obtained a maximum hydrogen yield of 3.1 mmol/g at 650 ºC. Wei *et al.* [25] experimentally pyrolyzed cellulose in molten carbonate obtaining a maximum hydrogen yield of 8.3 mmol/g of cellulose at 600 ºC. Although thermochemical pyrolysis generally depicts significantly higher hydrogen yield, typically of 22 to 128 mmol/g cellulose, as reported by [30], [58]–[60], the SDBD plasma process operates under significantly lower temperatures (< 200 ºC).

The yields of CO and $CO_2$ are the second and third largest, respectively, for both treatment times. This is attributed to the large amount of oxygen atoms present in cellulose. The higher yield of CO, slightly greater for $t_{treatment}$ = 15 min than for $t_{treatment}$ = 3 min, is caused by the partial oxidation reaction of cellulose [61]. Light hydrocarbons, i.e., $CH_4$, $C_2H_4$, and $C_2H_6$ are also produced by the process, but with significantly smaller yield. The overall yield trend of $H_2$ > CO > $CO_2$ observed in the SDBD plasma process has also been observed by Du *et al.* [61] during the gasification of corn cob via nonthermal plasma.

*Selectivity* quantifies the relative yield of a gas product compared to the total gas products and is presented in Fig. 8b. Hydrogen has the greatest selectivity of 76.7 and 88.1 vol.% for $t_{treatment}$ = 3 and 15 min, respectively. The greater hydrogen selectivity for $t_{treatment}$ = 15 min suggests that some of the lower molecular weight hydrocarbons are further decomposed into hydrogen and carbon. Several studies on the pyrolysis of biomass reported hydrogen selectivity. Du et al.[61] gasified corn cob using nonthermal plasma and reported hydrogen selectivity of about 60%. Wu et al. [58] catalytically pyrolyzed cellulose for hydrogen production using nickel-based catalysts (Ni-Zn-Al, 1:1) and obtained a selectivity of about 55 vol.%. Zsinka et al. [62] reported the highest hydrogen selectivity of 19% during the pyrolysis of biomass waste using modified nickel catalysts at 800 °C. Turn et al. [59] and Zeng et al. [26] reported hydrogen selectivity of 57 and 26 vol.%, respectively. The results obtained in the present work indicate that the treatment of cellulose by SDBD plasma can lead to greater hydrogen selectivity than thermochemical and thermo-catalytic approaches.

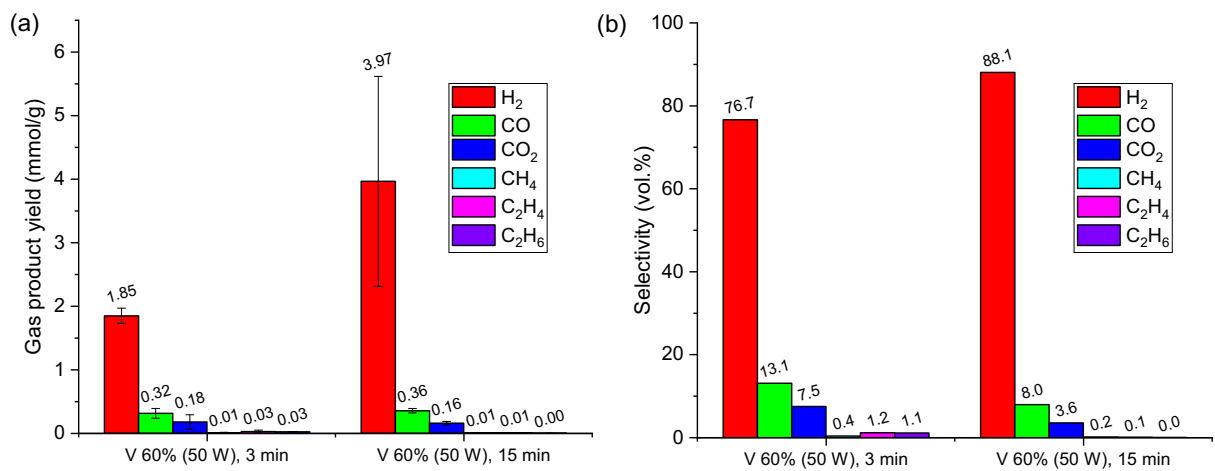

**Fig. 8. Gas product yield and selectivity.** (a) Yield and (b) selectivity of different gas products generated during the SDBD plasma treatment of cellulose.

***Solid product characterization.*** Images of the cellulose samples before and after plasma treatment as a function of time are presented in Fig. 9a. The pristine white sample ($t_{treatment}$ = 0 min) starts charring almost right from the start of the treatment and becomes more pronounced as the treatment time increases, transforming almost completely into char for a 15-minute treatment. The faster charring is likely attributed to the reactions involving the glycosidic oxygen atoms, leading to the formation of anhydrosugars which are subsequently converted into char and gas products [47], [48]. The non-uniform treatment of the sample is clearly observed at $t_{treatment}$ = 0.5 and 1 min, probably due to the porous nature of the cellulose feedstock (i.e., compacted powder). The porosity of cellulose creates less-resistive electrical paths leading to a localized concentration of discharge filaments. However, as the treatment time increases, this effect becomes less significant.

Results of CHN-elemental analysis of the samples (by Midwest Microlab) are presented in Fig. 9b. The carbon content and carbon-to-hydrogen weight ratio (C/H) of the solid sample increases with treatment time to a maximum of 85.6 wt.% (for $t_{treatment}$ = 6 min) and 26.6 (for $t_{treatment}$ = 15min), respectively. This indicates a significant extent of carbonization, as noted by Nanda et al. [63], who obtained 76.4 wt.% of carbon content in the catalytic gasification of wheat straw in hot compressed water for hydrogen production. The sudden variation of both carbon content and C/H indicates the non-uniformity of the plasma treatment of cellulose, which is also evidenced by the images in Fig. 9a.

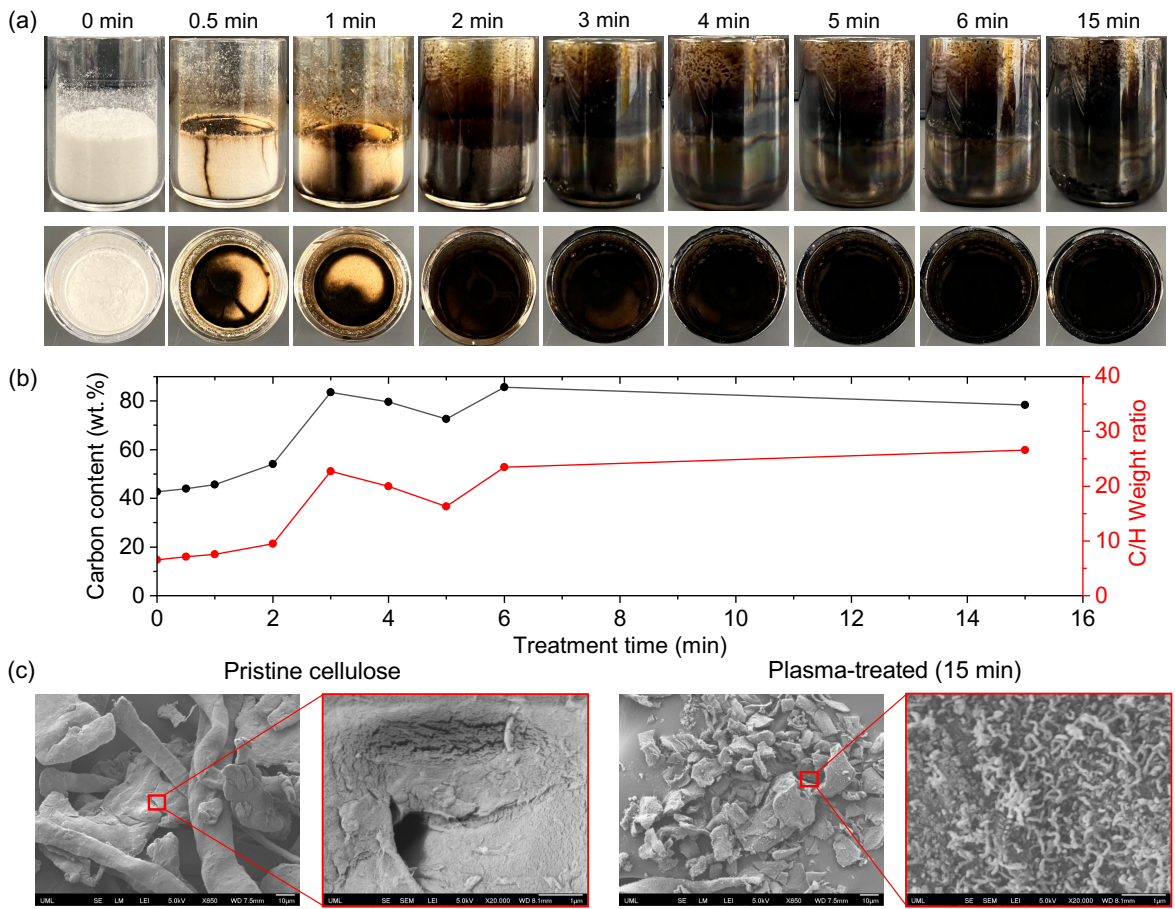

**Fig. 9. Characterization of treated cellulose feedstock.** (a) Optical images of cellulose samples before treatment ($t_{treatment}$ = 0 min) and at different treatment times. (b) Elemental characterization of cellulose as function of treatment time. (c) Surface morphological characteristics of pristine and plasma-treated cellulose samples obtained using FESEM under low- (850×) and high- (20000×) magnification.

Based on the optical observation of the solid residue with the greatest charring, the solid residue for $t_{treatment}$ = 15 min is selected for FESEM imaging together with the pristine sample. Representative imagining results are presented in Fig. 9c. The pristine sample consists of entanglements of cellulose fibers with empty spaces between fibers, as shown in the low-magnification images in Fig. 9c. The average width of the fibers is 100 $\mu$m. It is the crisscrossed network of fibers with empty spaces in-between fibers that lead to the high porosity of cellulose. The higher magnification image reveals that each fiber is made of bundled and indistinguishable fibrils, as observed by Du *et al.* [61]. The existence of pores creates the least resistive path for electric current, which probably accounts for the non-uniform treatment of the feedstock. Furthermore, the porous nature of cellulose is in part responsible for cellulose's weaker dielectric strength, which leads to lower power consumption and higher hydrogen production efficiency compared to that attained from LDPE's treatment (discussed in section 3.2).

The plasma-treated cellulose has fragmented fibers of small pieces, shown in Fig. 9c in the image under low magnification. The higher magnification image shows fragmented fibers that consist of protruded fibrils with an average diameter of 50 nm. These fibrils are loosely tangled and clearly visible and increase the surface area of the remnant solids. Du *et al.* [61] observed a similar structure in gasified corn cob using nonthermal plasma. Also, Zhang *et al*. [64] obtained carbon nanotubes of an average diameter of 50 nm using microwave-assisted chemical vapor deposition of carbon nanotubes on pine nutshell char.

### 3.2. Hydrogen production from low-density polyethylene

***SDBD plasma-feedstock interaction.*** The interaction between SDBD plasma and LDPE at the end of each treatment is depicted in Fig. 10. The sequence of images shows that initial violet emission from the plasma changes to yellow and eventually to red towards the end of 15 minutes of treatment. The yellow emission depicts the presence of oxygen attributed to the residual air in the reaction chamber and oxygen admixture during the pre-melting of the LDPE sample preparation. The presence of carbon particles is likely responsible for the red emission.

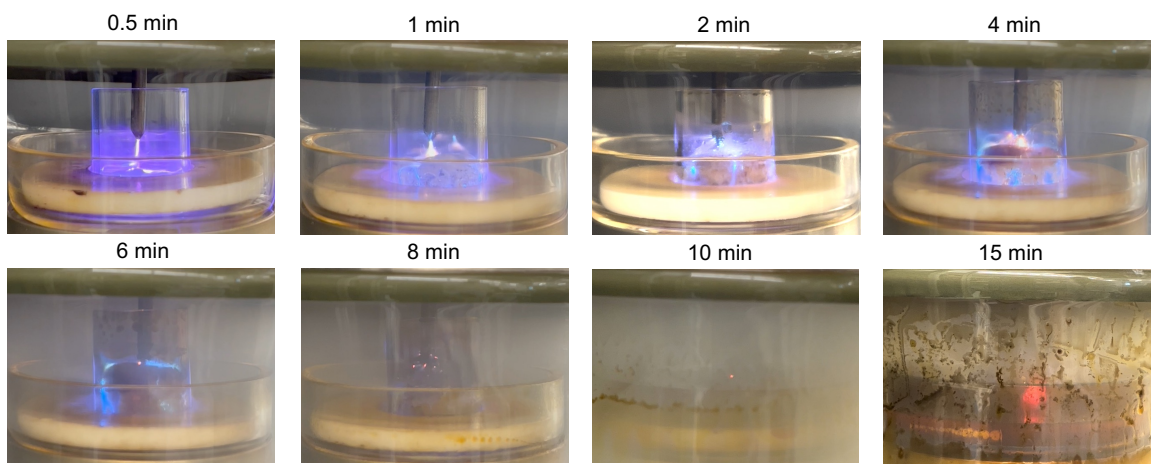

**Fig. 10. Optical imaging of plasma treatment of LDPE**. SDBD plasma interaction with LDPE for different times under operation conditions of $V = 60\%$ ($P_{t,rms} = 53$ W) and $Q = 0.01$ slpm.

The emission of particulate matter (smoke) observed after 1 minute of operation can probably be attributed to the beginning of charring/carbonization of the feedstock. Unlike cellulose, the intensity of the smoke during the treatment of LDPE remains relatively low in the first 6 minutes, depicting lower devolatilization. The presence of smoke is more pronounced as the treatment progresses. Gunasee *et al.* [65] noted that the devolatilization of cellulose occurs faster and at lower temperatures than that of LDPE. Such behavior is also observed in the present study. The slow devolatilization is probably attributed to the stronger carbon-hydrogen bond of 416.7 kJ/mol [66] exhibited by polyethylene. The results for $t = 10$ and 15 min reveal the presence of waxy deposits inside the reactor chamber, suggesting the formation of hydrocarbons as often observed in the pyrolysis of plastics.

The mechanism of decomposition of LDPE via plasma is still unclear, despite numerous studies conducted and growing interest [31]. It is well agreed that the highly energetic electrons and highly energetic heavy-species in plasma break the bonds between carbon-hydrogen (416.7 kJ/mol) and carbon-carbon (362.2 kJ/mol) [66], resulting in the production of free radicals and hydrocarbons [67], [68]. The formed free radicals, therefore, instigate the random chain scission [31]. Subsequently, hydrogen transfer generates secondary radicals and lightweight molecular hydrocarbons. It is also noted that further hydrocarbons are produced from the carbon-carbon scission by the primary radicals.

***Hydrogen production and production efficiency.*** Similarly, as done for the treatment of cellulose, the performance of the SDBD reactor to produce hydrogen from LDPE is assessed in terms of *Cumulative $H_2$ production*, *Mean $H_2$ production rate*, *$H_2$ Production efficiency*, and *Energy cost of $H_2$*. The definition of these metrics is similar as those presented in section 3.1. The *Cumulative $H_2$ production* of LDPE in Fig. 11a increases gradually with treatment time to a maximum of 3.1 mmol for $t_{treatment}$ = 15. Unlike cellulose, no distinct regime between devolatilization and char reaction is observed. The negligible hydrogen production at $t_{treatment}$ = 0.5 min suggests that the plasma power is consumed in melting LDPE. The lower amount of hydrogen produced from LDPE compared to that obtained from cellulose, given the same amount of feedstock used, can be expected given the stronger hydrogen bond within LDPE.

The *Mean $H_2$ production rate*, shown in Fig. 11b, indicates a rapid increase during the first 2 minutes of treatment to a peak rate of 20 mmol/h and after which it slightly declines. The slight decline in the production rate suggests a small depletion of the amount of hydrogen in the feedstock. The maximum mean hydrogen production rate is three times greater that obtained by Tabu *et al*. [32] in the experimental extraction of hydrogen LDPE via two different nonthermal plasma processes.

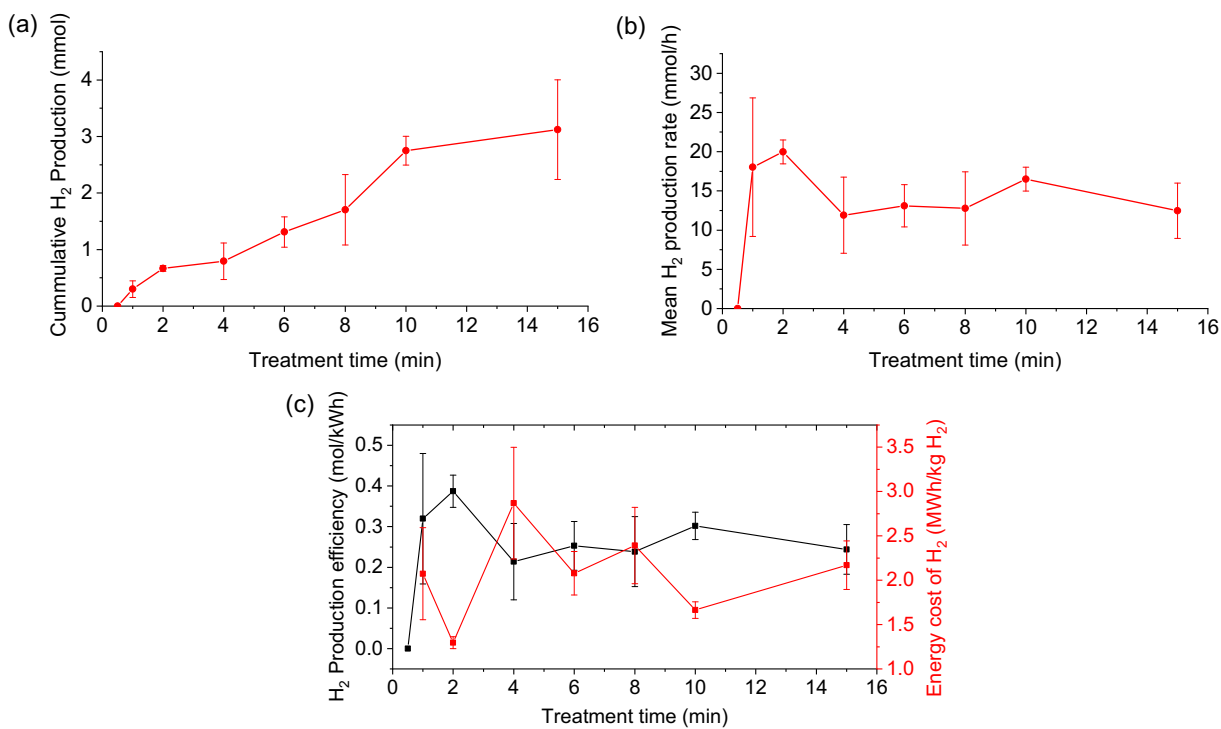

**Fig. 11. Hydrogen production performance of the SDBD plasma treatment of LDPE.** (a) Cumulative hydrogen production, (b) mean hydrogen production rate, (c) hydrogen production efficiency, and energy cost of hydrogen production as a function of treatment time.

The *$H_2$ Production efficiency* is often used as a metric to evaluate the viability of the process. Following the same trend as the *$H_2$ Production rate* (Fig. 11b), the hydrogen production efficiency, depicted in Fig. 11c (black line), increases rapidly in the first 2 minutes of treatment, before decreasing. The peak *$H_2$ Production efficiency* is 0.4 mol/kWh, half that obtained for cellulose. This is likely caused by the concerted reactions involving glycosidic oxygen atomics, leading to an enhanced decomposition of cellulose [47]. The decrease in production efficiency at the later time of treatment implies that the same amount of energy is consumed in extracting hydrogen from the hydrogen-deprived feedstock. The results also suggest that treating solidified pre-melted LDPE is undesirable as it leads to more compact nonporous feedstock with limited surface area. The smaller surface area limits the interaction between reactive species produced from the plasma and the hydrogen and carbon within LDPE, lowering hydrogen production.

The *Energy cost of $H_2$*, depicted in Fig. 11c (red line), represents the amount of energy required to produce one kilogram of hydrogen. As for cellulose, the *Energy cost of $H_2$* has an inverse relation with $H_2$ *Production efficiency*. It should be noted that the energy cost of hydrogen production for $t_{treatment}$ = 0.5 min is undefined since no quantifiable amount of hydrogen is detected by gas chromatography. The minimum energy cost obtained is 1300 kWh/ kg $H_2$, approximately two times greater than that of cellulose, and about 30 times greater than that for water electrolysis. Also, the energy cost is a 2-factor less expensive than the 3300 kWh/kg $H_2$ obtained in a previous study by the authors [32]. The high energy cost of hydrogen production is attributed to the stronger carbon-hydrogen bonds of LDPE.

***Gas product yield and selectivity.*** The *Gas product yield* and *Selectivity* are determined using the same definitions used for cellulose, namely equation (17) and equation (18), respectively. Yield and selectivity are determined for the hydrogen production experiments for $t_{treatment}$ = 4 and 15 min, and they are summarized in Fig. 12. The gas species identified and quantified are $H_2$, $CO$, $CO_2$, $CH_4$, $C_2H_4$, and $C_2H_6$. Given that LPDE does not contain oxygen, the presence of CO and $CO_2$ in the gas products is ascribed to residual air in the reactor and oxygen admixture during the pre-melting, a process of LDPE sample preparation (section 2.2). The maximum hydrogen yield of 3.12 mmol/g LDPE obtained in the 15-minute treatment of LDPE is comparable to what has been reported by other authors. Aminu *et al*. [29] studied the plasma catalytic steam reforming of high-density polyethylene (HDPE) and obtained hydrogen yield of about 4.5 mmol/g of HDPE. Nguyen and Carreon [31] reported hydrogen yield of 8 mmol/g of HDPE from a catalytic deconstruction of HDPE via nonthermal plasma. Alvarez *et al*. [19], in the catalytic pyrolysis of biomass and plastic, reported a hydrogen yield of 25.5 mmol/g HDPE.

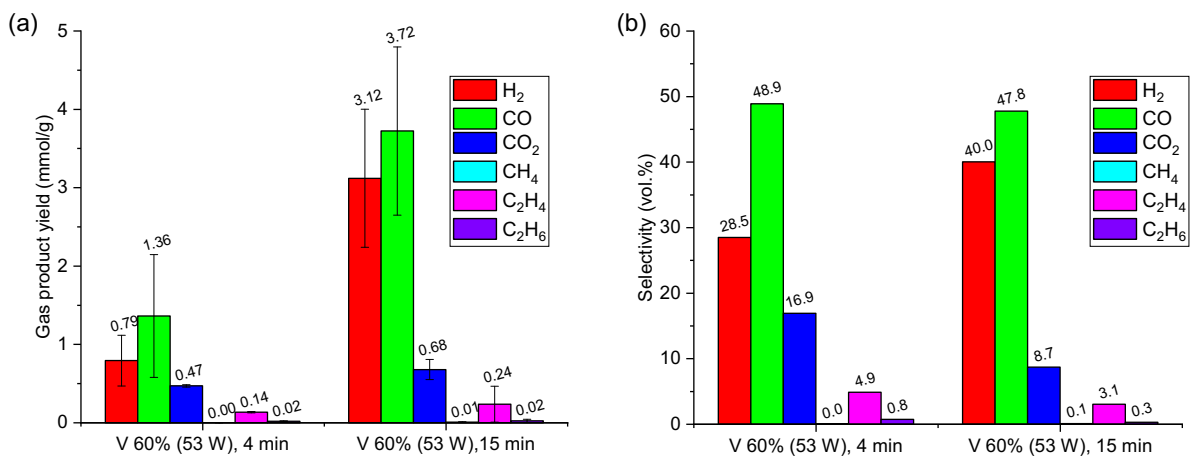

**Fig. 12. Gas product yield and selectivity.** (a) Yield and (b) selectivity of different gas products generated during the SDBD plasma treatment of LDPE.

The *Selectivity* of the different gas species produced is presented in Fig. 12b. The selectivity of hydrogen is 28.5 and 40.0 vol.% for $t_{treatment}$ = 4 and 15 min, respectively. The low hydrogen selectivity is attributed to the presence of residual air in the reactor chamber and oxygen admixture during the pre-melting of the LDPE sample leading to the formation of CO and $CO_2$. The highest yield of CO compared to $CO_2$ is probably attributed to the partial oxidation of LDPE in the presence of oxygen. The obtained selectivity is significantly lower than in other plasma-based processes. Nguyen and Carreon [31] obtained hydrogen selectivity of 50 vol.% in the catalytic deconstruction of HDPE via nonthermal plasma, whereas Farooq *et al*. [69] reported a hydrogen selectivity of 76 vol.% in the catalytic pyrolysis of LDPE.

*Solid product characterization.* Results of the characterization of the LDPE samples before and after SDBD plasma treatment is summarized in Fig. 13. As depicted in Fig. 13a, the white color of the pristine compact LDPE sample is transformed into brown as the treatment progresses and eventually to black for $t_{treatment}$ = 10 and 15 min. Contrary to cellulose, the plasma treatment of LDPE is more uniform, as depicted by an evenly-distributed browning of the sample for $t_{treatment}$ = 1 and 2 min. The solidified molten LDPE is denser and nonporous hence providing stronger dielectric resistance, which leads to the spreading of the plasma over the surface instead of leading to a localized electrical discharge. The light brown sample at $t_{treatment}$ = 1 min probably depicts the beginning of carbonization, whereas the blackening is attributed to charring.

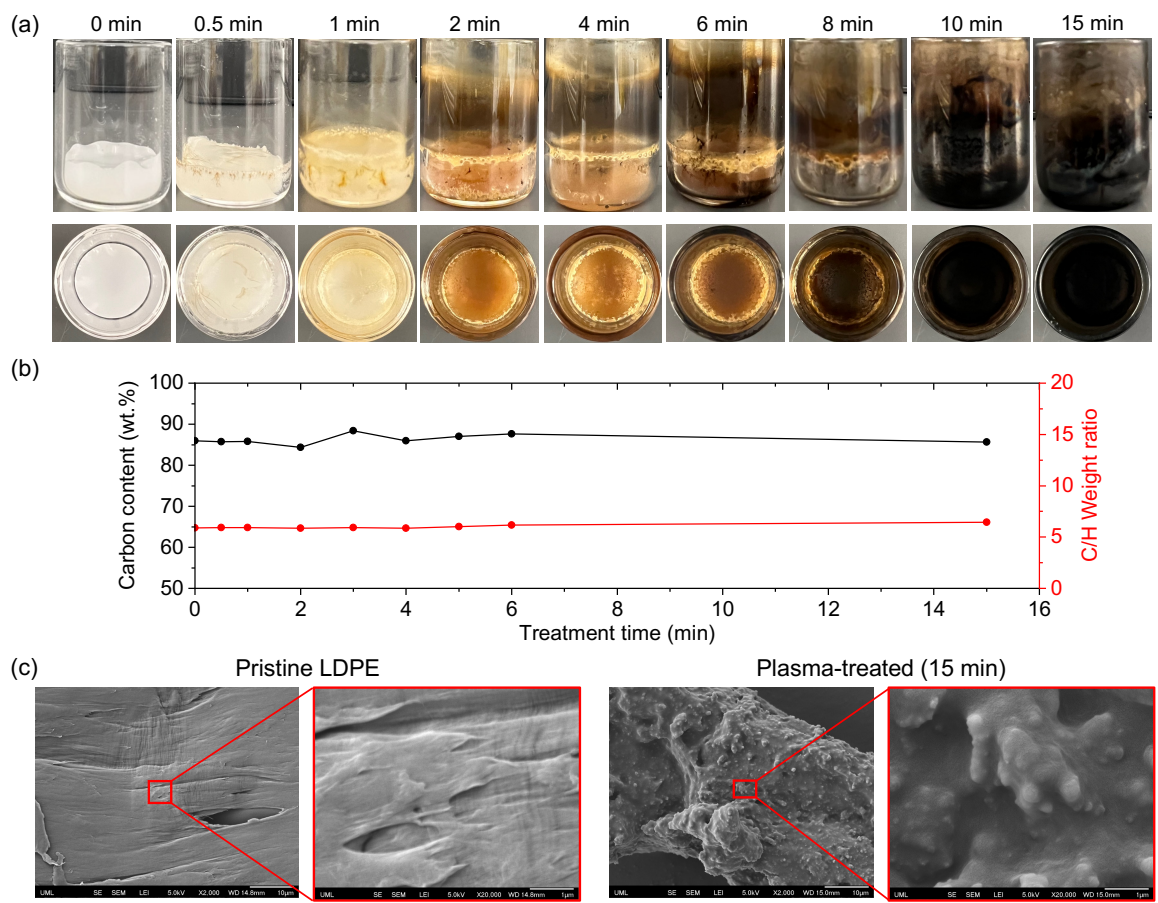

**Fig. 13. Characterization of treated LDPE feedstock.** (a) Optical images of LDPE samples before treatment (*t* = 0 min) and at different times. (b) Elemental characterization of LDPE as a function of treatment time. (c) Surface morphological characteristics of pristine and plasma-treated LDPE samples obtained using FESEM under low- (2000×) and high- (20000×) magnification.

The elemental characterization of the solids shows that the carbon content and C/H ratio vary minimally with treatment time, as shown in Fig. 13b. The unnoticeable increase in C/H depicts a very low degree of carbonization, and this suggests the treatment of solidified pre-melted LDPE is undesirable. The solidified pre-melted LDPE is associated with higher dielectric strength and limited reaction surface area, which require more energy and longer for a greater degree of carbonization.

The pristine LDPE morphology depicted by the FESEM result, presented in Fig. 13c, is a smooth and dense surface under low magnification with no observable pores. However, under high magnification, the surface is less smooth and non-uniform, with some fine peeling, which can be attributed to the mechanical damage during the sample extraction for the FESEM imaging. The denser and nonporous nature of LDPE shown by the microscope imaging appears consistent with its high dielectric strength, leading to a more uniform treatment compared to that for cellulose. Moreover, the dense and nonporous nature of LDPE also suggests greater power consumption during the plasma treatment (Fig. 11).

The microscopy images of the plasma-treated LDPE revealed substantive changes in the surface morphology shown in Fig. 13c. The surface is rough and embedded with micro-grains faintly visible under low magnification. It is interesting to note that, under higher magnification, the surface exhibited shallow dimples with micro-grains well embedded within the sample. This is probably attributed to the bombardment of the LDPE sample by the highly energetic electrons and ions during SDBD plasma treatment. The observed micro-grains are comparable to the carbon nanospheres observed by Kibria and Rashid [70] for the low-temperature synthesis of carbon nanomaterials, and those by Panickar *et al*. [71] in the chemical vapor deposition synthesis of carbon spheres, respectively.

## 4. Conclusions

The valorization of polymeric solid waste via low-temperature atmospheric pressure plasma could lead to economic and environmental benefits, particularly when powered by renewable electricity. In this study, a Streamer Dielectric-Barrier Discharge (SDBD) reactor is designed and built to extract hydrogen and carbon co-products from cellulose and low-density polyethylene (LDPE) as model feedstocks of biomass and plastic waste, respectively.

Experimental characterization and modeling indicate that the plasma consumes between 90 to 100% of the input power; the electron and excitation temperatures depict approximately the same value, independent of feedstock, and increase with input power up to ~ 1.6 to 2.4 eV; and the electron number density is ~ 2.5 $10^{13}$ cm$^{-3}$ irrespective of the feedstock and input power.

The mean hydrogen production rate for plasma-treated cellulose and LDPE increases to a maximum, after which it declines. The occurrence of peak production suggests that the plasma treatment of cellulose depicts two regimes, namely devolatilization and char reaction, similarly as observed in other biomass pyrolysis studies. The peak mean hydrogen production rate for cellulose is 40 mmol/h, which is twice that of LDPE (20 mmol/h). Moreover, the energy costs of hydrogen production for cellulose and LDPE are 600 and 1300 kWh/kg of $H_2$, respectively. The lower energy cost for cellulose is probably owed to its high porosity, leading to weaker dielectric strength that promotes increased hydrogen production at lower input power. Additionally, the hydrogen selectivity of cellulose is about two times more than that of LDPE due to the presence of residual gas in the reaction chamber and oxygen admixture during LDPE sample preparation, resulting in the production of CO and $CO_2$ in addition to $H_2$ and hydrocarbons.

The characterization of solid products via field emission scanning electron microscopy reveals distinct morphological structures of the two feedstocks. Whereas pristine cellulose comprises fibrils bundled into fibers with porous and entangled structures, pristine LDPE is a nonporous, uniform, and dense-structured compound. The plasma-treated cellulose consists of protrusions of an average diameter of 50 nm, while residual LDPE has embedded micro-grains, and both present promising valuable solid residues which need further investigation. The results indicate that the use of SDBD plasma is an effective approach for the production of hydrogen from cellulose and from LDPE at near atmospheric pressure and relatively low-temperature conditions in rapid-response and compact processes.

## 5. Acknowledgments

This work has been supported by the US Army Combat Capabilities Development Command (DEVCOM) Soldier Center Contracting Division through Contract W911QY-20-2-0005. The authors also acknowledge Danya Carla Maree for analyzing gas products from cellulose treatment.